\documentclass[aps,prb,twocolumn,groupedaddress,showpacs]{revtex4}
\usepackage{graphicx}
\usepackage{color}
\usepackage{amssymb,amsmath}
\usepackage{bm}
\def\tiw{\widetilde\omega}
\begin{document}
\title{Interplane and intraplane heat transport in quasi two-dimensional nodal
superconductors}
\author{I. Vekhter}
\author{A. Vorontsov}
\affiliation{Department of Physics and Astronomy,
             Louisiana State University, Baton Rouge, Louisiana, 70803, USA}
\date{\today}

\begin{abstract}
We analyze the behavior of the thermal conductivity in quasi-two
dimensional superconductors with line nodes. Motivated by
measurements of the anisotropy between the interplane and
intraplane thermal transport in CeIrIn$_5$ we show that a simple
model of the open Fermi surface with vertical line nodes is
insufficient to describe the data. We propose two possible
extensions of the model taking into account a) additional
modulation of the gap along the axial direction of the open Fermi
surface; and b) dependence of the interplane tunneling on the
direction of the in-plane momentum. We discuss the temperature
dependence of the thermal conductivity anisotropy and its low $T$
limit in these two models and compare the results with a model
with a horizontal line of nodes (``hybrid gap''). We discuss
possible relevance of each model for the symmetry of the order
parameter in CeIrIn$_5$, and suggest further experiments aimed at
clarifying the shape of the superconducting gap.

\end{abstract}
\maketitle

\section{Introduction}

Symmetry classification of possible gap structures established the
framework for separating conventional superconductors from their
unconventional counterparts
\cite{EBlount:1985,GVolovik:1985,MSigrist:RMP91}. Superconductors
with the order parameters that transform according to the trivial
irreducible representation of the point symmetry group of the
crystal are usually labeled conventional. If the order parameter
transforms according to a non-trivial representation of the same
group, the superconductor is called unconventional.

In the latter class the symmetry of the order parameter is lower
than the full group symmetry. This lowering of symmetry is usually
believed to be related to strong repulsive Coulomb interactions
and specific pairing mechanisms (for example, due to magnetic
fluctuations). As a result, determining the gap structure is one
of the crucial steps in testing our understanding of the origin of
superconductivity in a given material. An important group of
unconventional superconductors are those with order parameter that
vanishes (has nodes) for some directions at the Fermi surface.

Thermal conductivity is an exceptionally powerful probe for
testing the shape of the gap in nodal superconductors since only
the unpaired electrons (with momenta close to the nodal
directions) carry entropy.  The presence or absence of the nodes,
and sometimes their type (point or line), can be inferred from the
dependence of the thermal conductivity on the magnetic field and
temperature. An additional test for the existence of line nodes is
the so-called universality of the low-temperature coefficient of
the thermal conductivity, $a_0=\lim_{T\rightarrow 0}\kappa/T$
~\cite{PALee:1993,YSun:1995,MJGraf:1996}. The universal behavior
refers to the relative insensitivity of $a_0$ to the concentration
of the impurity atoms and to the details of the scattering on
individual impurity centers \cite{ACDurst:2000}. Physically, since
impurity scattering is pairbreaking, it generates near-nodal
quasiparticles which can carry heat, and whose lifetime is, in
turn, is limited by scattering on the same impurities. For linear
nodes the two effects cancel nearly exactly. In the high-T$_c$
cuprates, for example,  the quantitative agreement between
theoretical estimates of $a_0$ and the measured thermal
conductivity was established \cite{LTaillefer:1997,MChiao:2000}.
However, determining the location of the nodes in the momentum
space from such measurements is much harder.

The most direct tests measure the variation in the heat transport
with the orientation of the applied magnetic field with respect to
the nodal directions
\cite{FYu:1995,HAubin:1997,KIzawa:PrOsSb,KIzawa:YNiBCkappa,TWatanabe:UPd2Al3,KIzawa:CeCoIn5,KIzawa:BEDT}.
Long before such experiments were attempted, it was proposed  that
the {\it anisotropy} of the thermal conductivity along two
different directions as a function of temperature allows to infer
information about the gap structure
\cite{BArfi:1989,AFledder:1995}. Essentially, the measurement
determines the predominant direction of the Fermi velocity for the
nodal quasiparticles. Combined with the knowledge of the Fermi
surface, measurements of the evolution of the anisotropy in the
superconducting state impose stringent constraints on the possible
loci of the nodes.

This method was applied to UPt$_3$\cite{BLussier:1994}, and the
results argued convincingly for a line of nodes at $k_z=0$.  The
main reason this experiment did not uniquely determined the gap
structure is that theoretically expected results depend on the
detailed shape of the Fermi surface and the superconducting gap,
i.e. on the basis functions for the particular representation
\cite{MNorman:1996}. This detail apart, it is believed that the
measurement provides a good test for the horizontal vs. vertical
linear nodes.

Very recently the anisotropy of the thermal conductivity along
inequivalent crystalline directions was measured in heavy fermion
CeIrIn$_5$ by Shakeripour et al.~\cite{HShakeripour:2006}. The
ratio of the $c$-axis to the in-plane thermal conductivity,
$\kappa_c/\kappa_a$, rapidly decreases in the superconducting
state. Similarity of the anisotropy evolution to that for UPt$_3$,
where equatorial line of nodes is believed to exist, led  the
authors of Ref.~\onlinecite{HShakeripour:2006} to suggest that the
superconducting gap in CeIrIn$_5$ also has a horizontal line of
nodes. This is surprising since CeCoIn$_5$, a close relative of
CeIrIn$_5$, has vertical lines of nodes
\cite{KIzawa:CeCoIn5,HAoki:2004}, and the prevailing belief is
that the two compounds are quite similar (although some evidence
points to different origin of superconductivity in the two systems
\cite{MNicklas:2004}).

Consequently, in this paper we re-visit the analysis of the
temperature dependence and the anisotropy of the thermal
conductivity in nodal superconductors. Since both de Haas - van
Alphen measurements, and band structure calculations show that
CeIrIn$_5$ has an open, quasi-two dimensional piece of the Fermi
surface with significant $f$-electron contribution
\cite{YHaga:2001}, our main focus here is on such anisotropic
systems. We discuss the implications of the measurements for the
symmetry of the order parameter, and compare the anisotropy of the
thermal conductivity for several models relevant not only to
CeIrIn$_5$, but also to other systems with a quasi-two dimensional
Fermi surface.

The remainder of the paper is organized as follows. In next
section we introduce the basic experimental facts and theoretical
considerations, formulating the models. The subsequent sections
are devoted to the analysis of the proposed models. We then
critically compare our results with experiment, and propose
further measurements to test the gap symmetry in  CeIrIn$_5$.

\section{Experimental background}

CeIrIn$_5$ is a member of the so called 115 series, which also
include CeCoIn$_5$ and CeRhIn$_5$. While in the Rh system the
$f$-electrons of Ce remain localized and undergo antiferromagnetic
ordering, both Ir and Co compounds are paramagnetic heavy fermion
metals. In both systems de Haas-van Alphen measurements and the
band structure calculations indicate that a major sheet of the
Fermi surface is quasi-two dimensional, although the energy
dispersion along the $c$-axis is substantial, as evidenced by a
very moderate anisotropy ($\sim 2$) between the out of plane and
in-plane normal state transport coefficients. In contrast to
CeCoIn$_5$, which is a highly unconventional metal, likely close
to a quantum critical point at finite magnetic field
\cite{ABianchi:2003,JPaglione:2003}, CeIrIn$_5$ near the
superconducting phase is a good Fermi liquid, and therefore its
excitations should be adequately described by the  calculations in
the Fermi-liquid framework.

Both CeIrIn$_5$ and CeCoIn$_5$ are ambient pressure
superconductors, and Knight shift measurements indicate singlet
pairing \cite{YKohori:2001,GZheng:2001}. Specific heat and
in-plane thermal conductivity measurements
\cite{RMovshovich:2001}, as well as penetration depth
\cite{EChia:2003} and spin-lattice relaxation rate
\cite{YKohori:2001,GZheng:2001}, show the existence of line nodes.
In CeCoIn$_5$ there is overwhelming evidence for the vertical line
nodes from the anisotropy of the thermal conductivity and specific
heat under rotated magnetic field
\cite{KIzawa:CeCoIn5,HAoki:2004}, and from the tunneling into the
Andreev bound states \cite{WPark:2005}. Given the strong
similarities between the Fermi surfaces it is tempting to conclude
that the gap structure is similar in the two materials. There is,
however, evidence pointing towards differences in the origin (and
hence possibly in the type) of the superconducting state of the Co
and Ir compounds \cite{MNicklas:2004}.

Authors of Ref.\onlinecite{HShakeripour:2006} measured the
temperature dependence of the thermal conductivity in CeIrIn$_5$
along two inequivalent crystalline directions, $\kappa_{c}(T)$
(out of plane)  and $\kappa_{a}(T)$ (in plane), and made two
significant observations: a) The ratio $R(T)=\kappa_c/\kappa_a$ is
nearly temperature-independent above the superconducting
transition temperature, $T_c$, but is rapidly reduced with $T$ at
$T<0.5T_c$ (Note that inelastic scattering yields a peak in
thermal conductivity just below $T_c\sim 0.4$ K, which may lead to
the decrease of $R(T)$ appearing very pronounced); b) The low $T$
limit of the in-plane $\kappa_a/T$ appears universal, but the
interplane $\kappa_c/T$ does not  show the universal limit. Our
goal below is to examine the possible origin and implications of
this result.

\section{Basic considerations and models}

Independently of the choice of theoretical model, the results a)
and b) above imply, { prima facie},  that the effect of
superconductivity on the quasiparticles with the Fermi velocity
predominantly in the plane is different from that for the
quasiparticles moving along the c-axis: the states that carry
entropy along the $c$-axis have a larger ``effective gap'' than
those carrying heat current in the plane. Several processes of
different physical origin may lead to this behavior. In the
following we do not consider inelastic scattering; this is
appropriate at low temperatures, $T\ll T_c$. The range of
temperatures in Ref.~\onlinecite{HShakeripour:2006} where the
inelastic scattering can be neglected is not immediately clear
from the data, note, however, that, if the inelastic scattering is
isotropic, our conclusions regarding the anisotropy of the thermal
transport are not affected.

Since the Fermi surface of CeIrIn$_5$ has several sheets, it is,
of course, possible that the multiband effects are responsible for
the observed behavior. However, without a microscopic theory
detailing the gaps on each of the sheets, reaching reliable
conclusions about the agreement between theory and experiment
seems impossible, while phenomenological multiband  theory has too
many fitting parameters to seriously constrain possible gap
structures. We aim to construct a minimal model that captures the
essential physics of the the experimental observations, and
therefore restrict ourselves to considering one electronic band.

Since most band structure and dHvA analyses suggest an important
role of the quasi-two dimensional (open along the $c$-axis) band,
we choose such a band for our approach. We approximate it by a
simple nearest neighbor tight binding expression
\begin{equation}
  \epsilon(\bm k)=\frac{k_x^2+k_y^2}{2m}-2t(\bm k)\cos(k_zc),
  \label{Eq:energy}
\end{equation}
where $k_i$ are the components of the quasiparticle momentum,
${\bm k}$, $c$ is the lattice spacing in the $z$ direction, $m$ is
the effective mass, and $t(\bm k)$ is the interplane tunneling
matrix element. In all but one of our models we take $t(\bm k)$ to
be momentum-independent. We show below however, that the model
with $\bm k$-dependent interplane tunneling that provides a viable
path towards explaining the experimental results, even with the
assumption of vertical, rather than horizontal, nodes.

We consider singlet order parameters, and begin with the simple
model of vertical line nodes similar to the order parameter in
CeCoIn$_5$, introducing the azimuthal in-plane angle $\phi=\arctan
k_y/k_x$,
\begin{itemize}
  \item {\bf Model A:} $t(\bm k)=t$, and \\ \hspace*{1.6cm} $\Delta(\bm
  k)=\Delta_0\cos 2\phi$.
\end{itemize}
We show in the next section that this model is incompatible with
the experimental observations. Physically, the Fermi surface is
cylindrically symmetric and the gap depends only on the azimuthal
angle, $\phi$, while the $z$-axis component of the Fermi velocity
is $\phi$-independent. As we show below in Sec.~\ref{Sec:A}, these
conditions ensure that the ratio $\kappa_c/\kappa_a\equiv
\kappa_{zz}/\kappa_{xx}$, is temperature-independent even in the
superconducting state.

This constraint needs to be relaxed to explain the experimental
results, and we consider several possible models. First, we show
that for a tight-binding Fermi surface it can be expected, from
microscopic considerations, that the gap also acquires a weak
modulation along the $c$-direction, while retaining vertical lines
of nodes, and therefore consider
\begin{itemize}
  \item {\bf Model B:} $t(\bm k)=t$, and \\ \hspace*{1.6cm}
  $\Delta(\bm
  k)=\Delta_0(1+\delta\cos k_zc)\cos 2\phi$.
\end{itemize}
While in this model there exists a temperature-dependent
anisotropy, we find that this dependence is generally weak, and
hence unlikely to provide a satisfactory explanation of the
experimental observations (although we cannot exclude it).

In the second approach to lifting the constraints of model A we
consider a situation with vertical line nodes, but where the nodal
quasiparticles are less efficient than the normal electrons (on
average) in transporting heat along the $c$-axis compared to the
$xy$ plane. To model this we note that the measured and calculated
Fermi surface of CeIrIn$_5$ is not rotationally symmetric
\cite{YHaga:2001}, and that the energy dispersion of the electrons
along the $z$-axis clearly depends on the direction of the
in-plane momentum. This dependence is expected in the $D_{4h}$
symmetry, and $t(\bm k)$ is determined by the structure of the
overlapping wave functions for the quasi-two dimensional band. We
therefore introduce the angle dependence into the interplane
tunneling, preserving the crystal symmetry in the plane, and
propose
\begin{itemize}
  \item {\bf Model C:} $t(\bm k)=t_0+t_1\cos^2 2\phi$, and
  \\ \hspace*{1.6cm}$\Delta(\bm
  k)=\Delta_0\cos 2\phi$.
\end{itemize}
For $t_1>0$ the interplane tunneling is smaller for the nodal
quasiparticles than for antinodal ones, which ensures the
$T$-dependence of the anisotropy ratio. The situation is somewhat
reminiscent of that in the high-T$_c$ cuprates where, in the
absence of orthorhombic distortion, the interplane tunneling of
the nodal quasiparticles is suppressed, i.e. $t_0=0\ $
\cite{TXiang:1996}, and therefore the observed temperature
dependence of the Josephson plasma resonance frequency
\cite{MGaifullin:1999} and the $c$-axis penetration depth
\cite{CPanagopoulos:1997,AHosseini:1998} is different from that
suggested by the simple density of states power counting. We show
that such a model gives a significant temperature-dependent
anisotropy of the thermal conductivity.

Finally, we consider a model with horizontal, rather than vertical
line nodes, similar to the hybrid $E_{1g}$ gap proposed for
UPt$_3$, with broken time-reversal symmetry. The basis function of
the representation is described as $k_z(k_x+ik_y)$. For open Fermi
surface we require periodicity along the $z$-direction, and the
appropriate basis function is $(k_x+ik_y)\sin k_zc$. The
excitation spectrum and the thermal conductivity are only
sensitive to the gap amplitude, $|\Delta(\bm k)|=k_\perp|\sin
k_zc|$, where $k_\perp=(k_x^2+k_y^2)^{1/2}$. Since our
consideration of Model B shows (see the analysis below) that weak
modulation does not change the anisotropy qualitatively, we ignore
it, and consider
\begin{itemize}
  \item {\bf Model D:} $t(\bm k)=t$, and
  \\ \hspace*{1.6cm}$|\Delta(\bm
  k)|=\Delta_0|\sin k_z c|$.
\end{itemize}
We show that, this model also gives a substantially
temperature-dependent ratio $\kappa_{zz}/\kappa_{xx}$. Moreover,
we point out that from the data of
Ref.~\onlinecite{HShakeripour:2006} it is impossible to
distinguish Model C from Model D, and suggest further measurements
to probe the gap structure in CeIrIn$_5$.

We are now ready to consider each model in detail. In order to
make connection with experiment, we largely focus on the thermal
conductivity and its anisotropy
    \begin{equation}
      R(T)=\frac{\kappa_{zz}(T)}{\kappa_{xx}(T)}.
    \end{equation}
We begin by briefly looking at the normal state properties.

\section{Normal State}
\label{Sec:Normal}

\begin{figure}
  \includegraphics[width=1.05in]{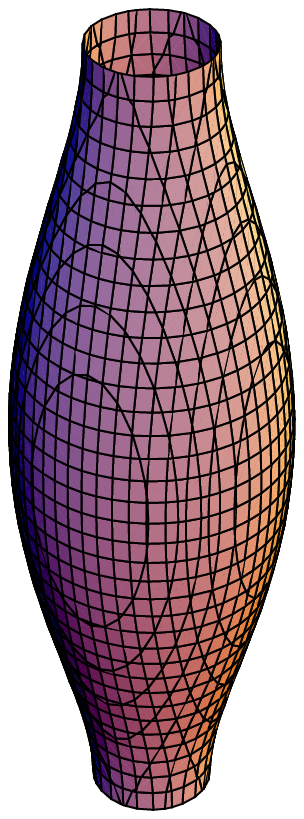}
  \hspace*{0.7cm}
  \includegraphics[width=1.2in]{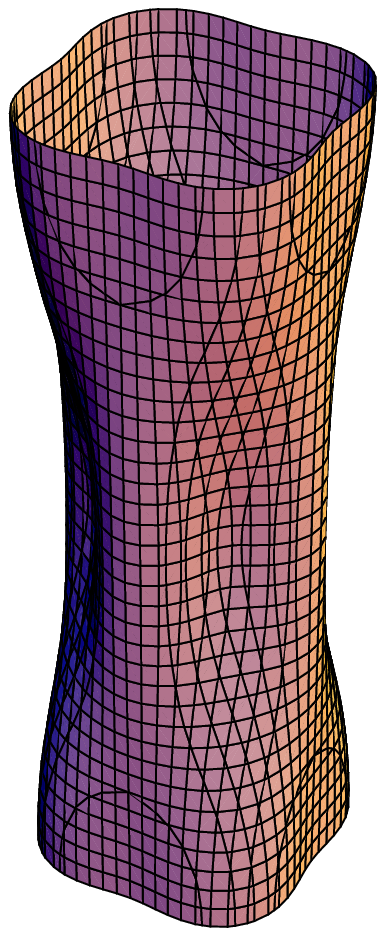}
  \caption{Fermi surfaces for models considered in the paper.
  Left panel: Models A,B, and D. Right panel: Model C. Notice the
  similarity between the Fermi surface shown in that panel and the Fermi surface
  suggested for CeIrIn$_5$
  in Ref.~\onlinecite{YHaga:2001}.}
  \label{Fig:FS}
\end{figure}

The Fermi surfaces for the models A through D are shown in
Fig.~\ref{Fig:FS}. In each case we transform to the cylindrical
coordinates and parameterize the Fermi surface by the azimuthal
angle $\phi=\arctan k_y/k_x$, and the c-axis quasimomentum $k_z$.
The Jacobian of the transformation from variables $k_x,k_y,k_z$ to
variables $\epsilon,k_z,\phi$ is unity, and therefore no
angle-dependence of the normal state density of states appears. We
denote the Fermi energy $E_F$, and define the Fermi momentum at
$k_z=\pm\pi/2c$ as $k_F=\sqrt{2mE_F}$.

In models A, B, D above the tunneling $t$ is $\phi$-independent.
In that case, the radius of the Fermi surface is
$k_0(k_z)=k_F\sqrt{1+(2t/E_F)\cos (k_z c)}$, and the components of
the Fermi velocity are
\begin{subequations}
\begin{eqnarray}
  v_x&=&\frac{k_0\cos\phi}{m}=v_F\sqrt{1+(2t/E_F)\cos (k_z c)}\cos\phi,
  \\
  v_y&=&\frac{k_0\cos\phi}{m}=v_F\sqrt{1+(2t/E_F)\cos (k_z c)}\sin\phi,
  \\
  v_z&=&2tc\sin (k_z c)=\frac{t}{E_F} (k_Fc)v_F\sin (k_z
  c),
\end{eqnarray}
\label{VF}
\end{subequations}
where $v_F=k_F/m$. In the normal
state the anisotropy of the transport coefficients is given simply
by
\begin{equation}
  R_n\equiv\frac{\sigma_{zz}}{\sigma_{xx}}=\frac{\kappa_{zz}/T}{\kappa_{xx}/T}=
  \frac{\langle v_z^2\rangle}{\langle v_x^2\rangle},
\end{equation}
where the average over the Fermi surface is
\begin{equation}
  \langle A\rangle=\int_0^{2\pi}\frac{d\phi}{2\pi}
  \int_{-\pi}^{\pi}\frac{d\chi_z}{2\pi}\, A(\phi,k_z),
  \mbox{ and } \chi_z=k_zc.
\end{equation}
For the simple model above $R_n=(t/E_F)^2(k_F c)^2$.

We can obtain a very rough estimate of the parameter values
relevant for CeIrIn$_5$ from the dHvA measurements and the band
structure calculations \cite{YHaga:2001}. For the
quasi-cylindrical piece of the Fermi surface the typical radius is
1/5 of the Brillouin Zone size, $k_F\approx 2\pi/(5a)$, where $a$
is the in-plane lattice constant. In this material $c/a\sim 1.6$,
so that $k_Fc \approx 2$. The normal state resistivity ratio,
$R_n\sim 0.4$, implying $t/E_F\sim 0.3$.

For model C the  situation is slightly more complex. We still have
\begin{equation}
  v_z=2t(\phi)c\sin (k_z c)=2t(\phi)c \sin \chi_z,
\end{equation}
and therefore
\begin{equation}
  \langle
  v_z^2\rangle=\frac{v_F^2}{2}(k_F c)^2\left[\frac{t_0^2}{E_F^2}+
    \frac{t_0t_1}{E_F^2}+\frac{3}{8}\frac{t_1^2}{E_F^2}\right].
\end{equation}
At arbitrary in-plane $k$ the $x$-component of the velocity is
\begin{equation}
  v_x=\frac{k\cos\phi}{m}+\frac{2t_1}{k}\sin\phi\sin
  4\phi\cos\chi_z,
\end{equation}
and, consequently, at the Fermi surface,
\begin{equation}
  \langle
  v_x^2\rangle=\frac{v_F^2}{2}+
    v_F^2 \frac{t_1^2}{E_F^2}\left<
    \frac{\sin^2\phi\sin^2
  4\phi\cos^2\chi_z}{1+2t(\phi)/E_F \cos\chi_z}
  \right>.
\end{equation}
Since we work in the regime $t_{0,1}/E_F\ll 1$, we can ignore the
corrections from the term $t(\phi)/E_F$ in the denominator as they
appear only at the order $(t/E_F)^4$, and write
\begin{equation}
  \langle
  v_x^2\rangle=\frac{v_F^2}{2}\left[1+\frac{3}{16}\frac{t_1^2}{E_F^2}\right].
\end{equation}
For the conductivity anisotropy of 2.5 cited above, the {\em
maximal} value of $t_1/E_F$, when there is no $t_0$ component, is
approximately $(t_1/E_F)^2\approx 4/15$, which corresponds to the
contribution of the second term of less than 5\% to the in-plane
transport in the normal state. Its effect in the superconducting
state is even smaller, since the $t_1$-dependent part of $v_x$
vanishes along the nodal directions. Consequently, we ignore the
$t_1$-dependent contribution to the in-plane transport hereafter.

\section{Model A: constant hopping, vertical line nodes}
\label{Sec:A}

The thermal conductivity of a superconductor with the gap function
whose average over the Fermi surface vanishes is given by
\cite{MNorman:1996,CKubert:1998}
\begin{eqnarray}
    &&\frac{\kappa_{ii}}{T} =\frac{N_0}{4}
    \int_0^{\infty}\frac{d\omega}{T}\left(\frac{\omega}{T} \right)^2
    {\rm sech}^2\left( \frac{\omega}{2T}\right) K_i(\omega , T),
\label{Eq:GenThermal}
\\
    \nonumber
       && K_i(\omega , T)
        =
    \frac{1}{\tiw_1 \tiw_2} \: {\rm Re} \left< v_i^2 \:
    \frac{\tiw^2 + |\tiw|^2 - 2|\Delta_{\bf k}|^2}
     {\sqrt{\tiw^2 - |\Delta_{\bf k}|^2}}
    \right>\; .
\end{eqnarray}
Here $N_0$ is the normal state density of states, and the
renormalized frequency $\tiw=\tiw_1+i\tiw_2=\omega-\Sigma(\tiw)$,
where $\Sigma$ is the self-energy due to impurity scattering. For
the model with $\Delta_{\bf k}=\Delta_0\cos 2\phi$, and the
hopping that is $\phi$-independent, we immediately find for
different orientations of the heat current
\begin{eqnarray}
\label{Eq:AThermal}
   &&\left< v_i^2 \:
    \frac{\tiw^2 + |\tiw|^2 - 2|\Delta_{\bf k}|^2}
     {\sqrt{\tiw^2 - |\Delta_{\bf k}|^2}}
    \right>
     =\frac{2\eta_i}{\pi}
     \\
     \nonumber
     &&\qquad\qquad\times\left[
       \tiw \: {\rm E} \left( \frac{\Delta_0}{\tiw} \right)
     + \frac{|\tiw|^2 - \tiw^2}{2\tiw} \:
       {\rm K} \left( \frac{\Delta_0}{\tiw} \right)
\right] \; ,
\end{eqnarray}
where $E$ and $K$ are the complete elliptic integrals, and
\begin{eqnarray}
  \eta_x=v_F^2,\qquad
  \eta_z=(2tc)^2.
\end{eqnarray}
Therefore the anisotropy of the thermal conductivity in the
superconducting state remains unchanged compared to the normal
state, $R_s=R_n=(t/E_F)^2(k_Fc)^2$. Moreover, in this model the
universal (nearly independent on impurity concentration) low
temperature limit, obtained from Eq.(\ref{Eq:AThermal}), is
\cite{PALee:1993,YSun:1995,MJGraf:1996,MNorman:1996}
\begin{equation}
  a_{0,i}=\lim_{T\rightarrow 0}\frac{\kappa_i}{T}=
  \frac{\pi}{6}\frac{N_0\eta_i}{\sqrt{\gamma^2+\Delta_0^2}}
  E\left(\frac{\Delta_0}{\sqrt{\gamma^2+\Delta_0^2}}\right),
  \label{Eq:KappaUniv}
\end{equation}
for both in-plane and out of plane directions. In
Eq.(\ref{Eq:KappaUniv}) $\gamma$ is the low-energy scattering
rate, $\widetilde\omega(\omega=0)=i\gamma$, and $E$ is the
complete elliptic integral. For clean superconductors,
$\gamma\ll\Delta_0$, and we have
\begin{equation}
  \frac{1}{\sqrt{\gamma^2+\Delta_0^2}}
  E\left(\frac{\Delta_0}{\sqrt{\gamma^2+\Delta_0^2}}\right)\approx
  \frac{1}{\Delta_0}+O\left(\frac{\gamma^2}{\Delta_0^2}\right),
\end{equation}
so that $a_{0,i}=\eta_i\pi N_0/(6\Delta_0)$, universal and
independent of the impurity scattering rate.

This constant and temperature independent ratio $R_s(T)$ is in
contradiction to the experimental measurements of
Ref.~\onlinecite{HShakeripour:2006}, and therefore model A does
not provide a satisfactory description of the superconducting
state of CeIrIn$_5$.

\section{Model B: modulated gap}
\label{Sec:B}

\subsection{Microscopic justification}

The symmetry classification does not uniquely determine the
functional form of the superconducting gap around the Fermi
surface. The gap can be given by any combination of the basis
functions transforming according to the chosen irreducible
representation. For example, in a cubic system, both a constant
gap and that varying as $k_x^4+k_y^4+k_z^4$ have the same symmetry
properties.

Of course, any $\bm k$-space variation of the gap beyond that
imposed by symmetry requirements costs condensation energy, and
therefore, all other things being equal, the simplest basis
functions often provide the most stable superconducting state. A
conventional superconductor with a spherical Fermi surface
strongly prefers a fully isotropic gap, while a strongly
anisotropic quasi two-dimensional system with a circular Fermi
surface and the dominant pairing in the $B_{1g}$ channel is likely
to have the gap with the $k_x^2-k_y^2$ momentum dependence. This
is the reason for using the familiar notation of $s$-wave for the
former case, and $d_{x^2-y^2}$ for the latter.

On the other hand, superconductivity is an instability of the
Fermi surface towards pairing, and therefore it is natural to
expect that in real materials with complex Fermi surfaces the gap
structure is also more complex.  The correct gap can only be
determined from the microscopic theory of superconductivity.  To
solve Eliashberg equations in superconductors with electron-phonon
mediated pairing, Allen introduced the Fermi surface harmonics
\cite{PBAllen:1976}, which have the symmetry of the Fermi surface,
are orthonormal, and are convenient basis functions for the
expansion of the superconducting pairing field.

In the absence of microscopic theory we use similar symmetry
arguments, and go beyond lowest order basis functions. For
simplicity, we continue to consider the $d_{x^2-y^2}$ gap
symmetry, $\Delta(\bm{\hat k})\propto \cos 2\phi$; all our
considerations apply equally well to other gaps with vertical line
nodes, such as $d_{xy}$. Choice of this symmetry determines the
irreducible representation of the crystal point group. In a system
with the Fermi surface open along the c-axis, we need to add to
this point group the invariance with respect to translations by a
reciprocal lattice vector along $z$. In the even (odd) channel the
basis functions for translation are $\cos nk_z c$ ($\sin nk_z c$),
with $n$ integer. Therefore quite generally the singlet
$d_{x^2-y^2}$ gap has the form
\begin{equation}
  \Delta(\bm{\hat k})=\sum_n \Delta_n \cos (nk_z c) \cos 2\phi.
\end{equation}
Similar interaction for $s$-wave pairing was considered by
Bulaevskii and Zyskin~\cite{LNBulaevskii:1990}, and for $d$-wave
symmetry by Rajagopal and Jha~\cite{AKRajagopal:1996,SSJha:1997}.
Note that since even powers of $k_z$, and therefore $\cos(nk_z
c)$, transform according to a trivial representation of the
tetragonal point group, the functions with different $n$ belong to
the same representation and can be mixed.

The degree of mixing is, of course, determined by the pairing
interaction. We consider a separable model,
\begin{equation}
  V(\bm{\hat k},\bm{\hat k^\prime})=\Phi(\bm{\hat k})\Phi(\bm{\hat
  k^\prime}),
\end{equation}
where
\begin{equation}
  \Phi(\bm{\hat k})=\sum_n V_n \cos (nk_z c) \cos 2\phi.
\end{equation}
To illustrate the behavior of the model, and to be consistent with
keeping only the nearest neighbor layer hopping in the energy
dispersion, Eq.(\ref{Eq:energy}), we truncate the expansion at
$n=1$, so that
\begin{widetext}
\begin{equation}
  V(\bm{\hat k},\bm{\hat
  k^\prime})=V_0\left[1+\lambda_1(\cos\chi_z+\cos\chi_z^\prime) +
  \lambda_2 \cos\chi_z\cos\chi_z^\prime\right]\cos 2\phi\cos
  2\phi^\prime.
  \label{Eq:VMod}
\end{equation}
\end{widetext}
 To the same accuracy the
gap is given by
\begin{equation}
  \Delta(\bm k)=\Delta_0\left[1+\delta\cos \chi_z\right]\cos
  2\phi,
  \label{Gapc-ab}
\end{equation}
and is determined from the self-consistency equation (for pure
superconductor in the weak-coupling limit)
\begin{equation}
  \Delta(\bm k)=\pi T\sum_{|\omega_n|<\Omega_0}
  \int_{-\pi}^\pi \frac{d\chi_z}{2\pi}
    \int_0^{2\pi}\frac{d\phi}{2\pi}
    \frac{\widetilde V(\bm{\hat k},\bm{\hat k^\prime})\Delta(\bm{\hat k^\prime})}
    {\sqrt{\omega_n^2+\Delta^2(\bm{\hat k^\prime})}}.
\end{equation}
Here $\widetilde V(\bm{\hat k},\bm{\hat k^\prime})=N_0V(\bm{\hat
k},\bm{\hat k^\prime})$, $\omega_n=\pi T (2n+1)$ are the fermionic
Matsubara frequencies, and $\Omega_0$ is the cutoff energy for the
pairing interaction.

Formally, the gap given by Eq.(\ref{Gapc-ab}) allows for a
horizontal line of nodes when $\delta>1$. This is, however,
physically unlikely. In the interaction $V(\bm{\hat k},\bm{\hat
k^\prime})$ above, $\lambda_2$ is the relative strength of the
pairing in neighboring planes compared to the in-plane pairing, so
that $|\lambda_2|<1$. Generally we also expect $\lambda_1\sim
t/E_F<1$ ~\cite{LNBulaevskii:1990}, so that the gap modulation
along the $c$-axis is insufficient to produce horizontal lines of
nodes. Inversely, if the interaction strengths in different
channels in Eq.(\ref{Eq:VMod}) are comparable, we cannot truncate
the expansion at the lowest order terms. Therefore, while in
principle the symmetry considerations allow the gap in
Eq.(\ref{Gapc-ab}) to have both horizontal and vertical nodes, the
model considered here can only be used if it yields small to
moderate values of $\delta$.

Consider first the linearized equations for the transition
temperature, $T_c$. Introducing dimensionless coupling constant
$g=N_0V_0$, we find that $T_c$ and $\delta$ satisfy
\begin{subequations}
\begin{eqnarray}
  1&=&\frac{g}{2}\left[1+\frac{\lambda_1\delta}{2}\right]
  \ln\frac{2\gamma_E\Omega_0}{\pi T_c},
  \\
  \delta&=&\frac{g}{2}\left[\lambda_1+\frac{\lambda_2\delta}{2}\right]
  \ln\frac{2\gamma_E\Omega_0}{\pi T_c}.
\end{eqnarray}
\label{LinearizedEq}
\end{subequations}
Here $\gamma_E\approx 0.58$ is the Euler's constant. Note that in
the absence of the $\lambda_1$ term in the pairing interaction the
equations for the temperatures of the transition into the states
$\Delta_0\cos 2\phi$ and $\Delta_1 \cos\chi_z\cos 2\phi$ decouple,
and each is reduced to the standard BCS expression,
$T_c^{(n)}=(2\gamma_E/\pi)\Omega_0\exp(-2/g_{eff}^{(n)})$, with
the effective coupling constants in the $\cos n\chi_z$ channels
$g$ and $\lambda_2 g/2$ for $n=0$ and $n=1$ respectively. For
$\lambda_2/2<1$ this implies that $T_c^{(0)}>T_c^{(1)}$, so that
the simple $\cos 2\phi$ gives the dominant order parameter.

For the generic case $\lambda_1\neq 0$, however, the two channels
are coupled, and the transition occurs directly into the c-axis
modulated state $\Delta=\Delta_0(1+\delta\cos\chi_z)\cos 2\phi$.
The transition temperature is
\begin{equation}
  T_c=\frac{2\gamma_E}{\pi}\Omega_0\exp(-2x_0/g),
\end{equation}
where
\begin{eqnarray}
  x_0&=&\frac{2}{1+\lambda_2/2+D^{1/2}},
    \\
  D&=&
  \left(1-\lambda_2/2\right)^2+2\lambda_1^2.
\end{eqnarray}
We assumed that the coupling $\lambda_2$ is not too repulsive. The
transition temperature depends on the magnitude, but not on the
sign of $\lambda_1$. On the other hand, the modulation,
$\delta=2(x_0^{-1}-1)/\lambda_1$, depends on the sign of
$\lambda_1$, and can be either positive or negative.

The modulation amplitude, $\delta$, is generally $T$-dependent.
Consider the self-consistency equation at $T<T_c$,
\begin{widetext}
    \begin{subequations}
  \begin{eqnarray}
    1=g\pi T\sum_{|\omega_n|<\Omega_0}
    \int_{-\pi}^\pi \frac{d\chi_z^\prime}{2\pi}
    \int_0^{2\pi}\frac{d\phi^\prime}{2\pi}
    \left[1+\lambda_1\cos\chi_z^\prime\right]
   \frac{\left[1+\delta\cos\chi_z^\prime\right]\cos^2 2\phi^\prime}
   {\sqrt{\omega_n^2+\Delta_0^2(1+\delta\cos\chi_z^\prime)^2\cos^2
   2\phi^\prime}}
   \\
   \delta=g\pi T\sum_{|\omega_n|<\Omega_0}
    \int_{-\pi}^\pi \frac{d\chi_z^\prime}{2\pi}
    \int_0^{2\pi}\frac{d\phi^\prime}{2\pi}
    \left[\lambda_1 +
             \lambda_2 \cos\chi_z^\prime\right]
   \frac{\left[1+\delta\cos\chi_z^\prime\right]\cos^2 2\phi^\prime}
   {\sqrt{\omega_n^2+\Delta_0^2(1+\delta\cos\chi_z^\prime)^2\cos^2
   2\phi^\prime}}
  \end{eqnarray}
  \label{self-cons1}
  \end{subequations}
\end{widetext}
In the linearized form considered above, all the terms in the
integrand linear in $\cos\chi_z^\prime$, vanish by symmetry. On
the other hand, below $T_c$, when $\Delta^2$ term in the
denominator is important, these terms give a finite contribution
to the gap equation. Moreover, in contrast to the conventional
``one-channel'' systems, the cutoff frequency $\Omega_0$ cannot be
removed from the self-consistency equations completely. As a
consequence, $\delta$ acquires temperature dependence.

\begin{figure}
  \includegraphics[angle=270, width=3.25in]{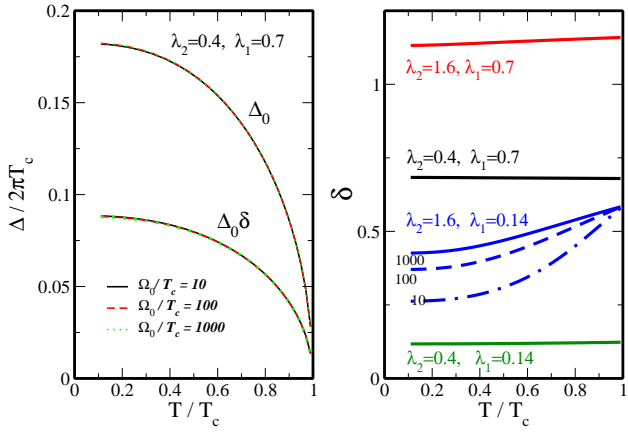}
  \caption{Temperatures dependence of the gap parameters in Model B.
  Left panel: temperature evolution of the gaps in two channels for typical
  coupling values. Note the independence of the result on the cutoff frequency
  $\Omega_0$: for different cutoffs $\Omega_0/T_c=10,100,1000$ the
  curves lie on top of one another.
  Right panel: Temperature dependence of the gap anisotropy, $\delta$ for different coupling
  constants. Note the temperature variation of the anisotropy in the regime of moderately large
  $\lambda_2=1.6$ and small $\lambda_1=0.14$, where we present the results for $\Omega_0/T_c=10,100,1000$. }
  \label{Fig:Del12}
\end{figure}

Of course, in the weak coupling limit  this dependence is
insignificant: in both integrals in Eq.(\ref{self-cons1}) the
greatest contribution is from the range
$\Delta_0\ll\omega_n\leq\Omega_0$, and therefore essentially
insensitive to the absence or presence of the gap. In most
situations the dependence on the cutoff vanishes already for
$\Omega_0/T_c\geq 10$. This is seen from the self-consistently
determined temperature evolution of the gaps in the two channels
computed using Eqs.(\ref{self-cons1}) and shown in the left panel
of Fig.\ref{Fig:Del12}. The corresponding plot of the modulation,
$\delta$, in the right panel of Fig.\ref{Fig:Del12} demonstrates
that the modulation is temperature-independent for most parameter
values.

We find that one exception is the somewhat artificial situation
when a) the bare transition temperatures into the two states are
very close, $\lambda_2\approx 2$ so that $T_c^{(0)}\approx
T_c^{(1)}$, {\em and} b) $\lambda_1\ll 1$. While $\delta$ is still
temperature-independent for $\Omega_0/T_c\rightarrow \infty$, the
temperature variations persist to large values of the cutoff, see
Fig.\ref{Fig:Del12}. Part of the reason is that the value of
$\delta(\lambda_1\rightarrow 0,\lambda_2\rightarrow 2)$ at $T_c$
obtained above depends on the order in which the limits are taken:
setting $\lambda_2=2$ first gives $\delta=\sqrt 2$, while taking
the limit $\lambda_1\rightarrow 0$ first for a finite value of
$1-\lambda_2/2$ yields $\delta=0$.

Physically, at the point $\lambda_2=2,\lambda_1=0$ the linearized
equations, Eq.(\ref{LinearizedEq}), decouple, and give no
information on the values of $\Delta_0$ and
$\Delta_1=\Delta_0\delta$. As is well known, the gap amplitudes
are determined by the fourth order terms in the Ginzburg-Landau
expansion of the free energy, or, equivalently, by the third order
terms in the gap equations. We find that inclusion of such terms
gives three different possible solutions for $\Delta_0$ and
$\Delta_1$. Two of them are trivial, $\Delta_0\neq 0,\Delta_1=0$
(lowest energy, as expected), and $\Delta_0=0,\Delta_1\neq 0$.
However, we find that there is a third solution with $\Delta_1/
\Delta_0=\delta=const$, which has the highest energy of the three.
At the same time for $\lambda_2=2$ and any {\em finite}
$\lambda_1$ $\delta=const$ corresponds to the lowest free energy,
which suggests a singular limit $\lambda_1\rightarrow 0$. This is
confirmed by carrying our the expansion of $\delta$ in $1-T/T_c$,
at finite $\lambda_1$, where we find that the coefficients are
singular for $\lambda_2=2$. This suggests a strong dependence of
the energies of the three stationary points on parameters, and on
temperature.

\begin{figure}
  \includegraphics[width=3.0in]{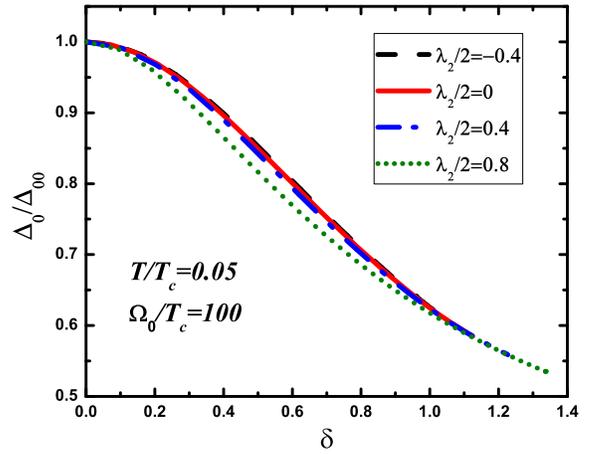}
  \caption{Gap amplitude as a function of the gap anisotropy for Model B at low
  temperature. The BCS value $\Delta_{00}=2.14k_BT_c$. Note the smooth evolution of the
  gap amplitude across the value $\delta=1$, where a horizontal line of nodes appears.}
  \label{Fig:Delta-delta}
\end{figure}

Away from the non-physical point $\lambda_2=2,\lambda_1=0$, the
numerical solution of the self-consistency equations,
Eqs.(\ref{self-cons1}) shown in Fig.\ref{Fig:Del12} makes it clear
that we can use the gap $\Delta=\Delta_0[1+\delta\cos\chi_z]\cos
2\phi$ with temperature independent $\delta$.  However, at low
temperatures $\Delta_0/T_c$ is no longer at the BCS ratio of 2.14,
but depends on the value of $\delta$. This dependence is computed
numerically, shown in Fig.\ref{Fig:Delta-delta}, and is used in
subsequent computation of the thermal conductivity.

\subsection{Thermal conductivity}

\begin{figure}[b]
  \includegraphics[height=2.5in]{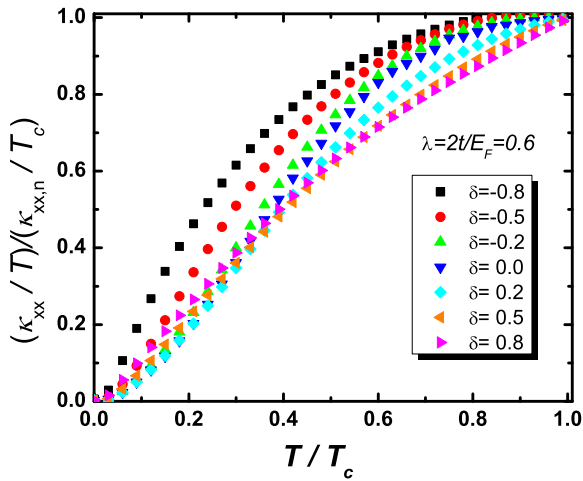}
  \caption{In-plane thermal conductivity for the modulated state.}
  \label{Fig:BKapx}
\end{figure}

\begin{figure}[t]
  \includegraphics[height=2.5in]{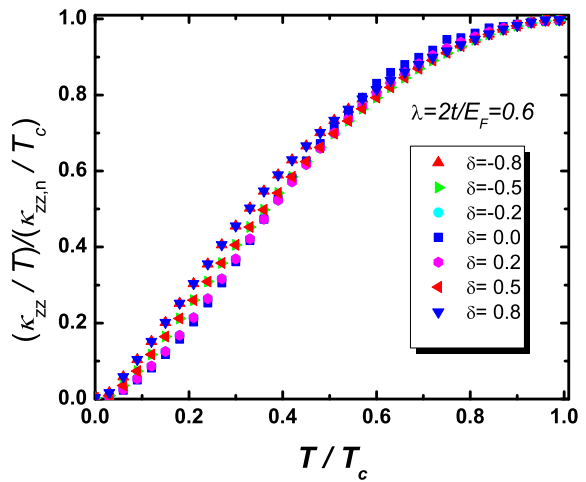}
  \caption{Interplane thermal conductivity for the modulated state.}
    \label{Fig:BKapz}
  \end{figure}

We use Eq.(\ref{Eq:GenThermal}) with
$\Delta=\Delta_0[1+\delta\cos\chi_z]\cos 2\phi$ to evaluate the
temperature dependence of the thermal conductivity in a quasi-two
dimensional superconductor. For our calculations we used the
impurity scattering in the unitarity limit, with the normal state
$\Gamma_N=0.007T_c$. While both $\kappa_{zz}$ and $\kappa_{xx}$
show the standard dependence on the scattering rate for
unconventional superconductors \cite{MJGraf:1996}, the anisotropy
ratio is essentially insensitive to impurity concentration in the
clean limit $\Gamma_N\ll T_c$. The results are shown in
Figs.~\ref{Fig:BKapx} and \ref{Fig:BKapz}. One point of note is
that the $c$-axis thermal conductivity is even in $\delta$, while
the in-plane transport is sensitive to whether the largest or the
smallest gap occurs for $k_z=0$, where the in-plane Fermi velocity
is the greatest, see Eq.(\ref{VF}).

\begin{figure}[t]
  \includegraphics[height=2.5in]{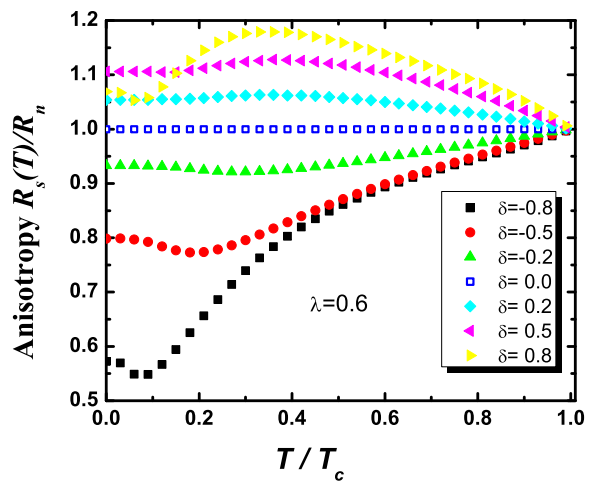}
  \caption{Anisotropy of the thermal conductivity for the model.}
  \label{Fig:BAnis}
  \end{figure}

Of immediate interest to us is the temperature dependence of the
anisotropy, $R(T)/R_n$, which is shown in Fig.~\ref{Fig:BAnis}.
The qualitative understanding of the temperature dependence relies
on the observation that the modulation changes the gap most
significantly in the regions $\chi_z=0,\pm\pi$, where the
quasiparticle velocity is strictly in the ab plane, with no
component along the $c$-axis. Consequently, for $\delta>0$ the
in-plane conductivity is reduced compared to that of a
superconductor with an unmodulated gap, while for $\delta<0$ it is
increased. As a result, for $\delta>0$ the anisotropy ratio
$\kappa_{zz}/\kappa_{xx}$ below $T_c$ is enhanced compared to the
normal state value, while for $\delta<0$ it is reduced. Notice
that for our model $\kappa_{zz}$ is insensitive to the sign of
$\delta$ since points at the Fermi surface where $\cos\chi_z=\pm
a$ ($\chi_z=\alpha,\pi-\alpha$) have identical values of the Fermi
velocity, and therefore contribute equally to the $c$-axis thermal
conductivity.

The residual anisotropy in the  $T\rightarrow 0$ limit can be
computed analytically once we recognize that, prior to integration
over the $z$-component of the momentum, for each value of $\chi_z$
the contribution of the kernel to the conductivity is universal in
complete analogy to a two-dimensional $d$-wave superconductor with
the gap $\Delta(\chi_z)$. Consequently, the ratio of the residual
low temperature terms is ($\lambda=2t/E_F$)
\begin{eqnarray}
\nonumber
  \frac{R_0}{R_n}&=& 2\left[\int_{-\pi}^{\pi}
                    \frac{\sin^2(\chi_z)d\chi_z}{1+\delta\cos\chi_z}\right]/
                        \left[\int_{-\pi}^{\pi}
                        \frac{(1+\lambda\cos
                        (\chi_z))d\chi_z}{1+\delta\cos\chi_z}\right]
    \\
    &=&
                        2\frac{1-\sqrt{1-\delta^2}}{\delta^2}
  \left[\frac{\lambda}{\delta}+\left(1-\frac{\lambda}{\delta}\right)\frac{1}{\sqrt{1-\delta^2}}
  \right]^{-1}.
\end{eqnarray}
This result is shown in Fig.~\ref{Fig:BAnis}. The experimentally
determined $R_s(T)/R_0$  is as low as 0.5 at temperatures of the
order of 0.2T$_c$ \cite{HShakeripour:2006}. This can only be
achieved for large negative values of the gap anisotropy,
$|\delta|=-\delta \gtrsim 0.8$, when the gap nearly vanishes in
the equatorial plane of the Fermi surface. This value is possible
for sufficiently strong coupling $\lambda_1$, but we consider it
not very likely in CeIrIn$_5$. In this model both $\kappa_{zz}$
and $\kappa_{xx}$ are universal, albeit with different values, and
therefore future doping studies of this material will be able to
test the universality of the $z$-axis thermal conductivity.

\section{Model C: modulated interplane hopping}
\label{Sec:C}

\subsection{Justification and general considerations}

In quasi-two dimensional systems the interplane matrix element
depends on the overlap of the atomic orbitals contributing to the
bands close to the Fermi surface. Therefore, in general, the
in-plane and out-of-plane motion of the quasiparticles are
coupled, and, in writing the tight-binding ansatz for the open
Fermi surface along the c-axis, it is necessary to recognize that
the matrix element $t$ is a function of the in-plane direction of
the momentum, $\phi$. Moreover, simply looking at the quasi-2D
sheet of the Fermi surface of CeIrIn$_5$ as obtained by
band-structure calculations and by dHvA
measurements~\cite{YHaga:2001}, it is clear that The Fermi surface
is not rotationally invariant, and therefore it is necessary to
account for the angle-dependence of tunneling. For the Fermi
surface of a general shape $k_F^2=k_x^2+k_y^2-2t\cos\chi_z$ the
difference between the maximal and minimal values of the radius,
$k_0^2(k_z)$ is $4t$. In the 115 series this difference is greater
along the [110] direction than along the [100] direction, which
suggest that $t$ is $\phi$-dependent~\cite{YHaga:2001}.

In the normal state this dependence does little beyond replacing
$t$ with the appropriate average over the directions, see
Sec.~\ref{Sec:Normal}. In a superconducting state with an
anisotropic gap, however, the effect of the modulation of the
interplane hopping can be much more pronounced. The best studied
example is the high-T$_c$ cuprates, where, on the basis of band
structure calculations, $t(\phi)$ is expected to vanish along the
directions of the nodes in the gap \cite{TXiang:1996}. Among the
manifestations of this effect are the high power law in the
temperature variation of the c-axis penetration depth ($T^5$
rather than $T$ in the pure limit) \cite{CPanagopoulos:1997}, and
weak dependence of the Josephson plasma resonance frequency on
temperature at $T\ll T_c$ ~\cite{MGaifullin:1999}.

In the analysis of the thermal conductivity it is easy to see that
any directional dependence $t(\phi)$ results in a deviation of the
anisotropy ratio, $R_s(T)$, from its normal state value, $R_n$,
even for a superconducting gap that has pure $d$-wave form,
$\Delta=\Delta_0\cos 2\phi$, with no $k_z$-dependence. This is
simply because the quasiparticle velocity, $v_z$, now acquires a
dependence on the angle $\phi$, which affects the evaluation of
the thermal conductivity kernel, $K_z$, see
Eq.(\ref{Eq:GenThermal}). Moreover, since the universal low
temperature behavior is due to near-nodal quasiparticles, any
suppression of the hopping matrix element in the vicinity of the
nodes reduces the contribution of these quasiparticles to the
c-axis transport, making it non-universal, while preserving the
universality of the low temperature limit for the in-plane
conductivity. Therefore assuming a modulated hopping may provide a
route towards the explanation of the experimental results.

\subsection{Model and thermal conductivity}

Once again we adopt the model energy dispersion of the form
\begin{equation}
    \epsilon(\bm k)=\frac{k_x^2+k_y^2}{2m}-2t(\phi)\cos(k_zc),
\end{equation}
where
\begin{equation}
  t(\phi)=t_0+t_1\cos^2 2\phi,
\end{equation}
which is the simplest anisotropic form satisfying the tetragonal
symmetry. This additional $\phi$ dependence does not affect the
in-plane thermal conductivity, $\kappa_{xx}$, but does modify the
out of plane $\kappa_{zz}$. As discussed in
Sec.{\ref{Sec:Normal}}, to a good approximation the normal state
anisotropy ratio is
\begin{equation}
  R_n=\frac{\kappa_{zz}}{\kappa_{xx}}=\left[\frac{t_0^2}{E_F^2}+
    \frac{t_0t_1}{E_F^2}+\frac{3}{8}\frac{t_1^2}{E_F^2}\right]
    (k_F c)^2.
\end{equation}
For simplicity, and in agreement with the results of the previous
section showing that small $c$-axis modulation of the gap does not
yield significant corrections to the anisotropy ratio in the
superconducting state, we consider a $d$-wave gap in this
calculation. In agreement with the discussion above, we are
interested in the situation when the interplane transport is
suppressed along the nodal directions, and therefore take
$\Delta=\Delta_0\cos 2\phi$.

Neglecting the small corrections to the in-plane Fermi velocity
due to the angle-dependence of $t$, as discussed in
Sec.\ref{Sec:Normal}, we find
\begin{equation}
  K_x=\frac{v_F^2}{\pi\tiw_1 \tiw_2} \: {\rm Re}
  \left[\tiw \: {\rm E} \left( \frac{\Delta_0}{\tiw} \right)
     + \frac{|\tiw|^2 - \tiw^2}{2\tiw} \:
       {\rm K} \left( \frac{\Delta_0}{\tiw} \right)
\right] \; .
\end{equation}
The kernel for the interplane conductivity is more complex,

\begin{widetext}
\begin{eqnarray}
  K_z=\frac{4c^2}{\pi\tiw_1 \tiw_2} \: {\rm Re}
  \Biggl\{
  t_0^2\left[\tiw \: {\rm E} \left( \frac{\Delta_0}{\tiw} \right)
     + \frac{|\tiw|^2 - \tiw^2}{2\tiw} \:
       {\rm K} \left( \frac{\Delta_0}{\tiw} \right)\right]
     +2t_0t_1\left[\tiw \: I_{1e} \left( \frac{\Delta_0}{\tiw} \right)
     + \frac{|\tiw|^2 - \tiw^2}{2\tiw} \:
       I_{1k} \left( \frac{\Delta_0}{\tiw} \right)\right]
     \\
      + t_1^2\left[\tiw \: I_{2e} \left( \frac{\Delta_0}{\tiw} \right)
     + \frac{|\tiw|^2 - \tiw^2}{2\tiw} \:
       I_{2k} \left( \frac{\Delta_0}{\tiw} \right)\right]
       \Biggr\},
\end{eqnarray}
where
\begin{eqnarray}
  I_{1e}(x)&=&\frac{2x^2-1}{3x^2} E(x)-\frac{x^2-1}{3x^2} K(x),
  \\
  I_{1k}(x)&=&\frac{K(x)-E(x)}{x^2},
  \\
  I_{2e}(x)&=&\frac{8x^4-3x^2-2}{15x^4}E(x)
            +\frac{2(1+x^2-2x^4)}{15x^4}K(x),
  \\
  I_{2k}(x)&=&-\frac{2}{3}\frac{1+x^2}{x^4}E(x)+\frac{2+x^2}{3x^4}K(x).
\end{eqnarray}
The $T=0$ limit, which is important for the comparison with the
universal limit, can also be readily evaluated as a function of
the low energy scattering rate, $\gamma$, by setting
$\widetilde\omega=i\gamma$. We reproduce the standard result for
the in-plane conductivity,
\begin{equation}
  \lim_{T\rightarrow 0}\frac{\kappa_{xx}}{T}
  =\frac{\pi}{6}\frac{N_0v_F^2}{\sqrt{\gamma^2+\Delta_0^2}}
  E\left(\frac{\Delta_0}{\sqrt{\gamma^2+\Delta_0^2}}\right),
\end{equation}
while for the interplane thermal conductivity we find
\begin{eqnarray}
  \lim_{T\rightarrow 0}\frac{\kappa_{zz}}{T}=
    \frac{2\pi}{3}\frac{N_0c^2}{\sqrt{\gamma^2+\Delta_0^2}}
    \Biggl\{&&
  t_0^2 E\left(\frac{\Delta_0}{\sqrt{\gamma^2+\Delta_0^2}}\right)
  +2t_0t_1\frac{\gamma^2}{\Delta_0^2}
  \left[K\left(\frac{\Delta_0}{\sqrt{\gamma^2+\Delta_0^2}}\right)
  -E\left(\frac{\Delta_0}{\sqrt{\gamma^2+\Delta_0^2}}\right)\right]
  \\
  &&
  +t_1^2\frac{\gamma^2}{\Delta_0^2}
  \left[
    E\left(\frac{\Delta_0}{\sqrt{\gamma^2+\Delta_0^2}}\right)
    -2\frac{\gamma^2}{\Delta_0^2}\left[
    K\left(\frac{\Delta_0}{\sqrt{\gamma^2+\Delta_0^2}}\right)
  -E\left(\frac{\Delta_0}{\sqrt{\gamma^2+\Delta_0^2}}\right)
  \right]
  \right]
  \Biggr\}.
\end{eqnarray}
\end{widetext}

\begin{figure}[b]
  \includegraphics[height=2.5in]{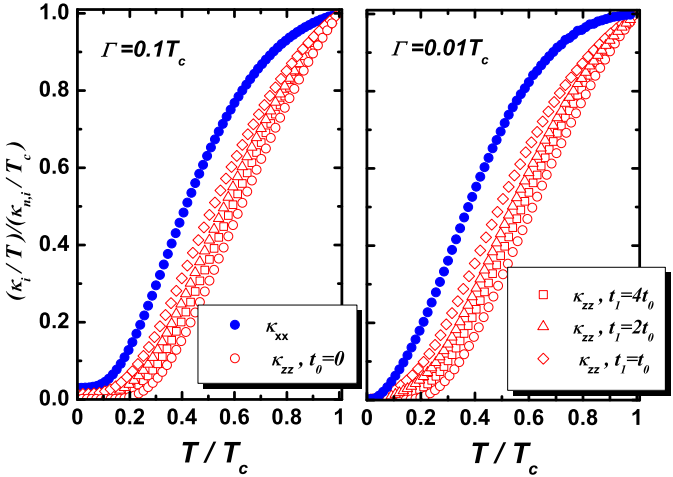}
  \label{Fig:CKap}
  \caption{Temperature dependence of the in-plane and
  the interplane thermal conductivity for the model with modulated
  hopping. The impurities are assumed to be in the unitarity limit,
  with $\Gamma$ the normal state scattering rate.}
\end{figure}

In the clean limit, $\gamma\ll \Delta_0$, the residual in-plane
conductivity is  universal,
\begin{equation}
  \lim_{T\rightarrow 0}\frac{\kappa_{xx}}{T}
  \approx\frac{\pi}{6}\frac{N_0v_F^2}{\Delta_0},
\end{equation}
while the $c$-axis conductivity, for $t_0/t_1>\gamma/\Delta_0$, is
given by
\begin{equation}
  \lim_{T\rightarrow 0}\frac{\kappa_{zz}}{T}\approx
    \frac{2\pi}{3}\frac{N_0c^2}{\Delta_0}t_0^2,
\end{equation}
which implies that the residual anisotropy ratio is
\begin{equation}
  \frac{R_0}{R_n}=\left[1+\frac{t_1}{t_0}+\frac{3t_1^2}{8t_0^2}\right]^{-1}<1.
\end{equation}
Even for a very moderate anisotropy ratio, $t_1/t_0=2$ (which, for
samples with $\gamma/\Delta_0\sim 0.1$ is well within the range of
the approximation), the anisotropy ratio at low temperature is
already low at about 22\%.

The reduction of the anisotropy is even more pronounced for the
cases of extreme anisotropy, $t_0/t_1\ll\gamma/\Delta$, when
\begin{equation}
  \frac{R_0}{R_n}\approx\frac{8}{3}\frac{\gamma^2}{\Delta_0^2}
  \left[1+\frac{t_0}{t_1}\ln\frac{4\Delta_0}{e\gamma}\right]\ll 1.
  \label{RAnis:C}
\end{equation}
Consequently in this model the residual anisotropy ratio can be
arbitrarily small depending on the purity. If there are no
symmetry requirements for the hopping matrix element to vanish for
the nodal directions, in most situations we do not expect $t_0$
and $t_1$ to differ by an order of magnitude; this suggests that
the residual linear term is visible in the $c$-axis thermal
conductivity, but may be significantly smaller than its in-plane
counterpart. While the results above give a rough estimate of the
effect of impurities on the residual anisotropy, it is worth
noting that disorder is likely to make the interplane hopping more
isotropic, and therefore tends to restore  the anisotropy to the
the normal state value faster that what Eq.(\ref{RAnis:C})
indicates.

\begin{figure}
  \includegraphics[height=2.5in]{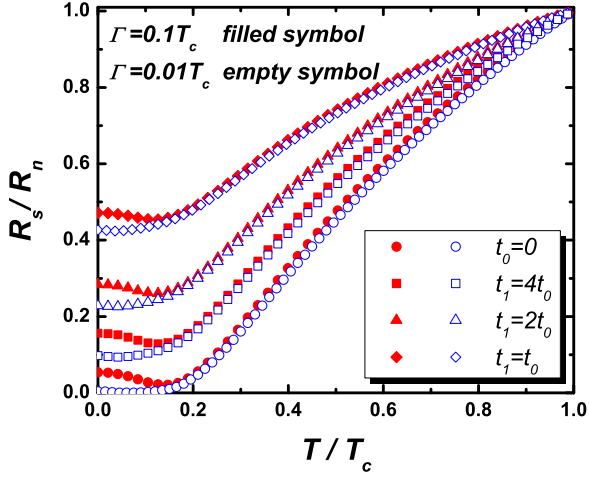}
  \label{Fig:CRs}
  \caption{Temperature dependence of the anisotropy of the
  interplane to the intraplane thermal conductivity. Note that the residual value is
  increasing with increasing $t_0/t_1$.}
\end{figure}

\section{Model D: hybrid gap}

So far we looked at the models with vertical line nodes, motivated
by possible similarities between CeIrIn$_5$ and CeCoIn$_5$, and
attempted to reconcile them with the experimental measurements. Of
course, in the absence of any information about the nodal
structure, a natural explanation for the observed anisotropy is
that, in analogy to UPt$_3$, the system has a horizontal line of
nodes, and therefore the nodal quasiparticles do not contribute to
the $c$-axis transport. We consider this model now.

\begin{figure}
  \includegraphics[height=2.5in]{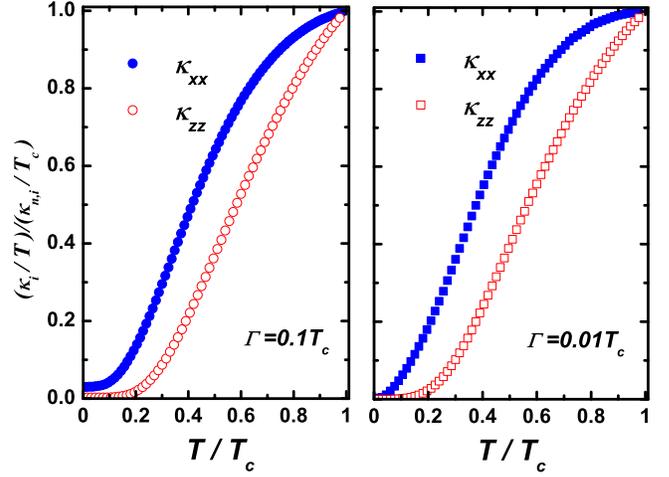}
  \label{Fig:DKap}
  \caption{Temperature dependence of the in-plane and
  the interplane thermal conductivity for the hybrid gap. The impurities are assumed to be in the unitarity limit,
  with $\Gamma$ the normal state scattering rate.}
\end{figure}

\begin{figure}
  \includegraphics[height=2.5in]{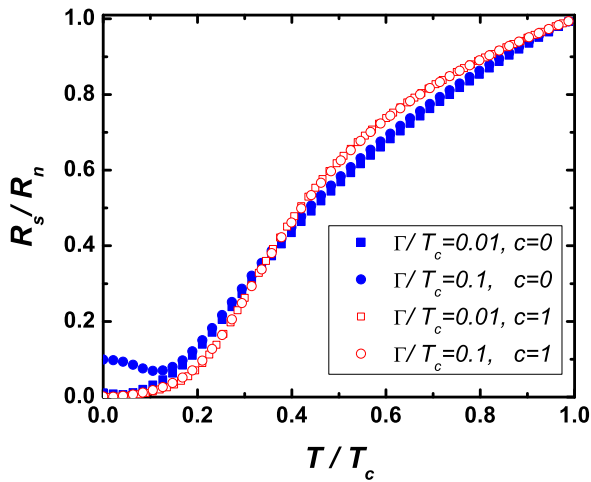}
  \label{Fig:DRs}
  \caption{Temperature dependence of the anisotropy of the
  interplane to the intraplane thermal conductivity for the hybrid gap. For comparison,
  we include results with deviations from the unitarity scattering limit.}
\end{figure}

The hybrid gap belongs to the representation that transforms as
$\widehat k_z(\widehat k_x\pm i\widehat k_y)$. The basis function
for this representation over an open Fermi surface is $(k_x\pm i
k_y)\sin\chi_z$, and therefore, for, if we take the weakly
modulated limit, $k_{F,x}^2+k_{F,y}^2\approx$const, the gap
amplitude is $|\Delta(\chi_z,\phi)|=\Delta_0|\sin\chi_z |$. If we
take into account the fourfold, rather than cylindrical, shape of
the Fermi surface, there may be additional small modulation of
this gap with the component of the Fermi momentum in the $xy$
plane as a function of $z$: based on our results for Model B in
Sec.\ref{Sec:B}, we ignore these. Then the density of states, and
the in-plane thermal conductivity for this model are identical to
that for a system with vertical line nodes, while the interplane
thermal conductivity kernel is given by
\begin{equation}
  K_z=\frac{4t^2c^2}{\pi\tiw_1 \tiw_2} \: {\rm Re}
  \left[\tiw \: I_{1e} \left( \frac{\Delta_0}{\tiw} \right)
     + \frac{|\tiw|^2 - \tiw^2}{2\tiw} \:
       I_{1k} \left( \frac{\Delta_0}{\tiw} \right)\right]
\end{equation}
The temperature dependence of $\kappa_{xx}$ and $\kappa_{zz}$ is
shown in Fig.~\ref{Fig:DKap}, and the corresponding anisotropy
ratio for the hybrid gap is given in Fig.~\ref{Fig:DRs} for the
hybrid gap for different values of the impurity scattering.  At
low $T$ the in-plane thermal conductivity is universal, while the
interplane conductivity,
\begin{eqnarray}
   \lim_{T\rightarrow 0}\frac{\kappa_{zz}}{T}&\approx&
    \frac{2\pi}{3}\frac{N_0t^2c^2}{\sqrt{\gamma^2+\Delta_0^2}}
    \\
    \nonumber
    &&\times
    \frac{\gamma^2}{\Delta_0^2}
    \left[
    K\left(\frac{\Delta_0}{\sqrt{\gamma^2+\Delta_0^2}}\right)
    -E\left(\frac{\Delta_0}{\sqrt{\gamma^2+\Delta_0^2}}\right)
    \right]
    \\
    &\approx&
    \frac{2\pi}{3}\frac{N_0t^2c^2}{\Delta_0}\frac{\gamma^2}{\Delta_0^2}
    \ln\frac{4\Delta_0}{e\gamma},
\end{eqnarray}
gives the residual anisotropy
\begin{equation}
  \frac{R_0}{R_n}\approx\frac{\gamma^2}{\Delta_0^2}
    \ln\frac{4\Delta_0}{e\gamma}.
\end{equation}
This result is clearly non-universal. Notice that the dependence
on the impurity scattering here differs from the $\gamma/\Delta_0$
result obtained for UPt$_3$ in Ref.~\onlinecite{MNorman:1996}. The
reason for the discrepancy is that for the hybrid gap and a Fermi
surface closed along the $c$-axis the quasiparticles contributing
the most to the $c$-axis conductivity are near the north and south
pole. For the open Fermi surface these quasiparticles are absent,
and the $c$-axis conductivity is reduced by an additional factor
of $\gamma/\Delta_0$. It is also obvious, from comparing this with
the results of the previous section, that the residual anisotropy
can be close for the models C and D, and may not provide the
unequivocal distinction between the two.

\section{Discussion.}

\begin{figure}
  \includegraphics[height=2.5in]{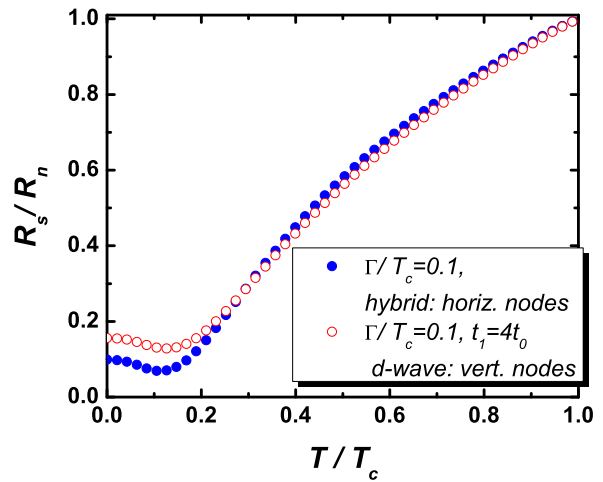}
  \caption{Comparison of the temperature dependence for the hybrid and the
  d-wave gap.}
  \label{Fig:CompRs}
\end{figure}

It follows from experiment that the low-$T$ value of the
anisotropy of the thermal conductivity is below 25\% of the normal
value. The analysis above shows that model A (simple $d$-wave with
constant interplane coupling) does not give results that can
explain the observed anisotropy. Model B (weakly modulated, along
the $c$-axis, amplitude of the $d$-wave gap) shows some tendency
towards the temperature-dependent anisotropy, but, in our view,
requires fine tuning if it were to explain the measurements of
Ref.~\onlinecite{HShakeripour:2006}. On the other hand, both model
C (modulated interlayer hopping, vertical line nodes) and model D
(horizontal line nodes) give very similar behavior of the thermal
conductivity as a function of temperature, with a low residual
value of the anisotropy $\kappa_c/\kappa_a$, and therefore may be
relevant to CeIrIn$_5$. Distinguishing between the two based
purely on the thermal conductivity data may not be easy, as is
seen from the comparison in Fig.\ref{Fig:CompRs}, which shows
that, for sufficiently high anisotropy of $t_1/t_0$, there is
essentially no difference in the behavior of the ratio
$\kappa_c/\kappa_a$ as a function of temperature between the two
models.

There are notable differences between the results presented for
either model and the experimental observations of
Ref.~\onlinecite{HShakeripour:2006}. In experiment, the anisotropy
ratio does not decrease below 0.5 for temperatures as low as
0.1$T_c$, and there is no sign of saturation of the ratio below
0.15$T_c$ as in theoretical analysis. This is due to strong
inelastic scattering that results in a peak in $\kappa_c/\kappa_a$
at about $0.3T_c$, but which is not included in our analysis here.
Correspondingly, the low-$T$ fit to experiment is done just below
the peak, and the rapid decrease of the anisotropy may be related
to the inelastic scattering.

The low residual value of the thermal conductivity anisotropy
implies, in model C, highly anisotropic tunneling, $t_1\geq 2t_0$.
Published band structure data do not allow to reliably extract
this ratio from fitting the quasi-two dimensional sheet of the
Fermi surface, but this is the task that perhaps should be
attempted in near future. In the meantime, we propose several
experiments that have the potential to provide additional
information to resolve the question of vertical vs. horizontal
line nodes.
\begin{enumerate}
  \item {\em Direct measurement of the specific heat and/or thermal
  conductivity in the vortex state under a rotated applied field.}
  This is, in our view, the best test for the
  existence of the line nodes. If, for the field rotated in the $xy$ plane,
   no difference in the specific heat
  is found across
  the $H$-$T$ phase diagram for the field direction along [110]
  vs. [100] direction, it is likely that there no vertical lines
  of nodes. If, on the other hand, such anisotropy is found and
  develops in agreement with theoretical predictions
  \cite{IVekhter:1999,AVorontsov:2006}, this would be a strong
  evidence for $d$-wave, rather than hybrid, gap.

  At the same time, it is possible, although, in our view,
  unlikely, that there is gap modulation {\em both} in the plane
  and along the $c$-axis. Therefore measurements of the specific
  heat and thermal conductivity when the field is rotated in the
  $zy$ or $xz$ plane should be used to eliminate this
  possibility, and directly probe for horizontal line nodes. Low
  superconducting transition temperature of CeIrIn$_5$ and inelastic scattering
   require
  that these  measurements be done at temperatures below $\sim$150mK.

  \item {\em Measurement of the $c$-axis thermal conductivity in
  CeCoIn$_5$.} In the Co compound there is a general agreement that
  there are vertical line nodes. Measurements of the temperature evolution
  of the anisotropy
  $\kappa_c/\kappa_a$ in that compound, and comparison with
  CeIrIn$_5$, while short of proof, would provide a test for the connection between
  the anisotropy and the location of the nodes. Note that there is some
  evidence for
  remaining unpaired electrons in the superconducting state of
  CeCoIn$_5$ (at least upon La-doping) \cite{MATanatar:2005}, which leads to a residual
  metallic linear term for both directions of the heat current.
  However, as it was suggested that the fraction of these quasiparticles
  can be
  determined experimentally \cite{MATanatar:2005}, their
  contribution can be subtracted, and the remaining anisotropy
  used to test the differences and similarities between CeCoIn$_5$
  and CeIrIn$_5$.

  \item {\em $c$-axis penetration depth measurements.} In analogy
  to cuprates, if the interplane transport is suppressed along the
  nodal directions, the $c$-axis penetration depth should either
  be not linear in $T$ at $T\ll T_c$ (for $t_1\gg t_0$) or have a
  small range of linear $T$ behavior, with the coefficient much
  smaller than that inferred from the density of states varying as
  $E/\Delta_0$. In CeCoIn$_5$ the $c$-axis penetration depth is
  linear in $T$, and observation of the non-linear behavior in
  CeIrIn$_5$ would point towards differences between the two
  systems. Unfortunately, we expect that, for reasons similar to
  those outlines for the thermal conductivity, it would be
  difficult to distinguish model C from model D on the basis of
  this measurement. Negative result, however (linear, in $T$,
  $\lambda_c(T)$) would be difficult to reconcile with the thermal
  conductivity measurements.

  \item {\em Doping studies of residual $c$-axis thermal
  conductivity in CeIrIn$_5$.} In Fig.\ref{Fig:ResKap} we show the
  evolution of the residual thermal conductivity along the
  $c$-axis, in units of the expected universal value for that
  direction, as a function of the low-energy scattering rate,
  $\gamma$. Notice that the increase in $\kappa_{zz}/T$ for the
  hybrid gap is slow, while for the modulated interplane hopping
  of model C it is faster. In reality we expect that addition of
  impurities will tend to make the interplane hopping more
  isotropic in addition to introducing pairbreaking, and therefore
  the increase in the residual value of $\kappa_{zz}/T$ will be even
  faster than that suggested by Fig.\ref{Fig:ResKap}. We therefore
  predict that, if model C of vertical line nodes with modulated
  $t$ is realized, disordering the sample will produce a much more
  pronounced effect on the interplane thermal conductivity than
  that suggested by a hybrid gap.
  \begin{figure}
  \includegraphics[height=2.5in]{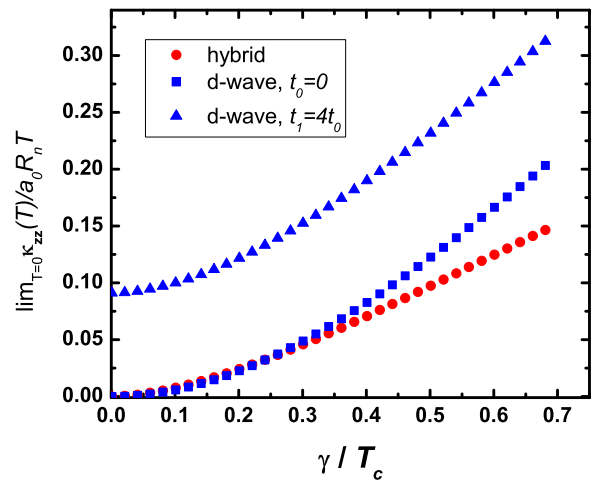}
  \caption{Increase of the residual thermal conductivity with impurity scattering rate.}
  \label{Fig:ResKap}
\end{figure}
\end{enumerate}

\section{Conclusions}

In conclusion, we investigated the constraints placed on the shape
of the superconducting gap in CeIrIn$_5$ by the results of
Shakeripour et al \cite{HShakeripour:2006}, assuming a one-band
model with the Fermi surface open along the $c$-axis of the
crystal. We find that the temperature evolution of the anisotropy
between the transport along the $c$-axis and in the plane is
incompatible with a simple $d$-wave gap over a quasi-two
dimensional Fermi surface, such as often assumed in the studies of
related compound, CeCoIn$_5$. The temperature variation of the
anisotropy shows that the $c$-axis conductivity is affected more
significantly than its intraplane counterpart by the opening of
the superconducting gap. This implies that the ability of the
unpaired quasiparticles to carry heat current along the $c$-axis
is impaired relative to their contribution to the in-plane
transport. This disparity can arise either from the gap anisotropy
or the Fermi surface properties. We explored several models where
such modulation has different physical origin.

We find that a weakly $c$-axis modulated $d$-wave gap does not
easily provide satisfactory agreement with experiment, although
may do so with fine tuning of the parameters. Of the models with
constant interplane hopping, that with {\bf horizontal line nodes}
most naturally account for the anisotropy, in partial agreement
with the argument of Shakeripour et al.

Remarkably, we also find that an alternative model, where the
$c$-axis hopping depends on the direction of the in-plane
quasiparticle momentum, yield results that agree with the
experimental data even if purely $d$-wave gap, with {\bf vertical
line nodes}, is assumed.  Such an agreement requires substantial
variation of the hopping between the nodal and the antinodal
directions, in qualitative agreement with the analysis of the
Fermi surface obtained in the dHvA measurements.

Given that there is some evidence for a distinct superconducting
dome in CeIrIn$_5$ it is clearly very important to determine the
shape of the gap.  We therefore suggested several possible
experiments aimed directly at distinguishing the two situations
and hope that further analysis will help resolve the uncertainty
regarding the gap symmetry in this material. We also believe that
the analysis above is generally relevant to a number of
unconventional superconductors with a quasi-two-dimensional parts
of the Fermi surface.

\section{Acknowledgements}

We are very indebted to H.~Shakeripour and L.~Taillefer for
discussing their results with us before publication, and to
M.~A.~Tanatar for correspondence.  We are also grateful to
C.~Capan for critical reading of the manuscript.  This research
was supported in part by the Board of Regents of Louisiana.

\end{document}